## Title

*In Vitro* and *In Silico* Characterization of the Aggregation of Thrombi on Ventricular Assist Device Cannula

## Authors

Wenxuan He[1], Abhishek Karmakar[2], Grant Rowlands[2], Samuel Schirmacher[2], Rodrigo Méndez-Rojano[2], James F. Antaki[2]

[1]Sibley School of Mechanical and Aerospace Engineering, Cornell University, Ithaca, NY, USA 14850

orcid=0000-0002-3255-5057

[2]Meinig School of Biomedical Engineering, Cornell University, Ithaca, NY, USA 14850

## The corresponding author

Dr. James F. Antaki, Meinig School of Biomedical Engineering, Cornell University, Ithaca, NY, USA 14850,

antaki@cornell.edu

## Graphic abstract

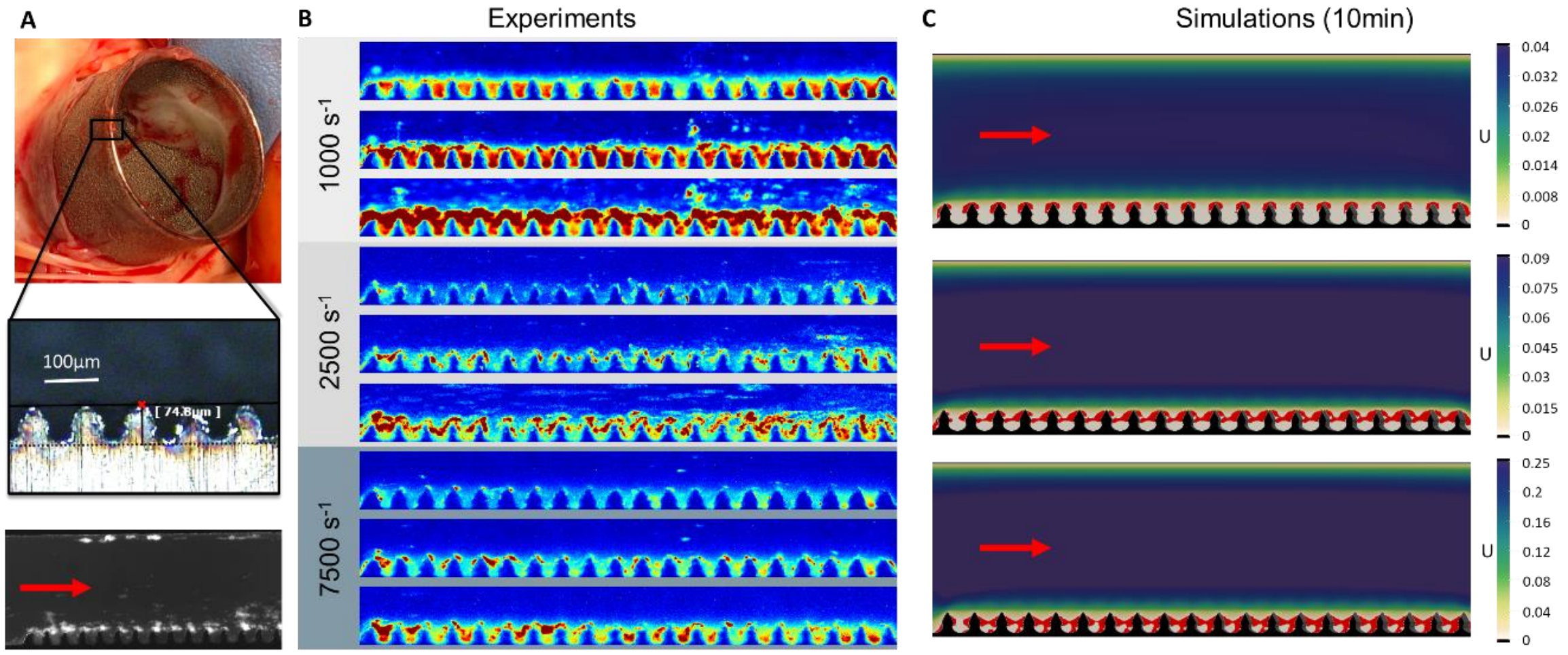

**Abstract**

Background

The unacceptably high stroke rate of HeartMate III VAD without signs of adherent pump thrombosis is hypothesized to be the result of the thrombi originating on the inflow cannula, ingesting and ejecting emboli from the VAD. Therefore, inflow cannula thrombosis has been an emerging focus. The inflow cannula of contemporary VADs, which incorporate both polished and rough regions serve as useful benchmarks to study the effects of roughness and shear on thrombogenesis. An *in vitro* study was conducted to emulate the micro-hemodynamic condition on a sintered inflow cannula, and to observe the deposition and detachment patterns. Together with a computational fluid dynamic tool, this study aimed to provide insight into the optimization of inflow cannula and potentially reducing adverse neurological events due to upstream thrombus.

Methods

A custom parallel plate channel was designed for this study which incorporates periodic teeth (approximately 75-micron Ra) that mimic the sintered surface treatment of a left ventricular inflow cannula. Anti-coagulated whole human blood (3.2% sodium citrate), collected from 17 healthy donors, was perfused through the microchannel by a syringe pump (Harvard Apparatus, Holliston, MA USA) at calibrated flow rates to achieve shear rates of 1000, 2500, and 7500$s^{-1}$ (Reynolds numbers 3.06 - 22.88). The deposition of fluorescently labeled platelets was visualized using inverted epifluorescence microscopy. Mean fluorescence intensity, deposition probability, and embolization pattern were collected. Numerical simulations of a previously validated thrombosis model were performed using OpenFOAM to simulate blood flow in the channel with periodic teeth. A hexahedral dominant mesh was used with the transition to prism elements near the refinement region at the boundary.

Results

The average fluorescence intensity map depicts the time-course deposition within the serrated micro-channel for three perfusion rates studied. The sustained growth of platelet adhesion with time was seen in all shear conditions. Increasing the shear rate results in less adhesion and more bulk embolization, which was captured by real-time video. Variance in deposition location was also found associated with an increasing shear rate, confirmed by CFD simulation. At the lowest shear rate, platelet aggregates developed onto the tips of the teeth and were trapped within the valleys between teeth. At 7500 $s^{-1}$, less deposition was found on the tips, and emboli were observed washing off from the tips.

Conclusions and Relevance

This *in vitro* and *in silico* thrombus formation study is the first to focus on emulating clinical-related surface roughness and supraphysiological shear of a VAD inflow cannula. The simulation reproduced the experimental observation of the shear-dependent spatial distribution of platelet deposition. The findings in the variance of adherent thrombus and mean intensity of platelet deposition may suggest that a supraphysiological shear results in a greater chance of embolization around the rough cannula surface, which raises concerns about hemocompatibility of inflow cannula without signs of adherent thrombus.

## Introduction

For nearly a decade, thrombosis has been a major hindrance in the universal adoption of ventricular assist devices (VADs) for advanced heart failure patients [1,2]. Furthermore, the thrombus is a precursor for adverse neurological events, one of which, thrombotic stroke, is a leading cause of morbidity and mortality in LVAD patients [3]. In June 2021, Medtronic (Fridley, MN USA) stopped the sale of the HVAD due to, in part, elevated stroke risks presented to the patient [4,5]. Although the occurrence of pump thrombosis has been radically reduced in HeartMate III ( $\leqslant$ 1.4% at 2 years post-implantation) [6], the incidence of ischemic stroke (10%) remains a concern [6]. The unacceptably high stroke rates without the sign of adherent pump thrombosis may result from thrombi originating on the ventricular inflow cannula that is ingested by the VAD and ejected as emboli [6–10]. (See **Error! Reference source not found.**A.) Therefore, inflow cannula thrombosis has been an emerging focus.

In an attempt to improve hemocompatibility manufacturers have intentionally roughening the surface of the inflow cannula to promote neointima or pseudoneointima [11–14]. For example, sintered microspheres were applied to the ventricular cannula of the HeartMate – I (aka XVE) and later adopted by others including the HeartMate II, HVAD, and HeartMate III, to encourage endocardial tissue to adhere at the insertion site and prevent embolization [14]. (See **Error! Reference source not found.**B and 1C.) Clinical experience with HVAD revealed that the ring of discontinuity between smooth and sintered regions was a nidus for thrombotic deposition. (See Figure 1E.) In contrast, the HeartMate 3 inflow cannula has sintered titanium on all blood-contacting surfaces: both interior and exterior to promote a confluent protective biological layer. (See **Error! Reference source not found.**C.) Nevertheless, there is inadequate evidence of the *genesis* of intended endothelialization – particularly in the early stages after implantation.

It is known that the initial response by blood to artificial biomaterial is characterized by platelet deposition, owing to the combination of surface chemistry, blood flow patterns, and coagulability [15]. Numerical studies of ventricular cannula have shown that supraphysiological shear (approximately 7000$s^{-1}$) may activate platelets and trigger deposition on the rim of the cannula [16,17]. However, the effect of the extreme surface roughness provided by titanium sintering on the accumulation and stability of thrombus is not well known. Prior *in vitro* studies of platelet deposition on roughened Ti6Al4V on the micron scale at elevated shear rate as high as 5500 $s^{-1}$ [18] has revealed a synergistic effect of shear rate and surface roughness to thromboembolism. Jamiolkowski et al. who studied platelet aggregation, adhesion, and embolization on individual crevices on the scale of sintered titanium found a negative correlation between platelet adhesion to Ti6Al4V and shear rate ranging from 400 $s^{-1}$ to 1000 $s^{-1}$ [19].

The present microfluidic experiments reported here were conducted to emulate the sintered surface of the inflow cannula to study the dynamics of platelet deposition and adhesion in real time under various supraphysiological shear conditions. The sintered topography was modeled as a series of crevices 70 µm in depth at a shear rate of 1000 - 7000 $s^{-1}$ based on literature data [14,16]. A complementary *in silico* study was conducted using a numerical model of thrombosis to simulate the time course of deposition. The overall aim was to provide insight into the platelet response to the sintered surface of the inflow cannula, potentially reducing adverse neurological events due to cannula thrombus.

## Materials and Methods

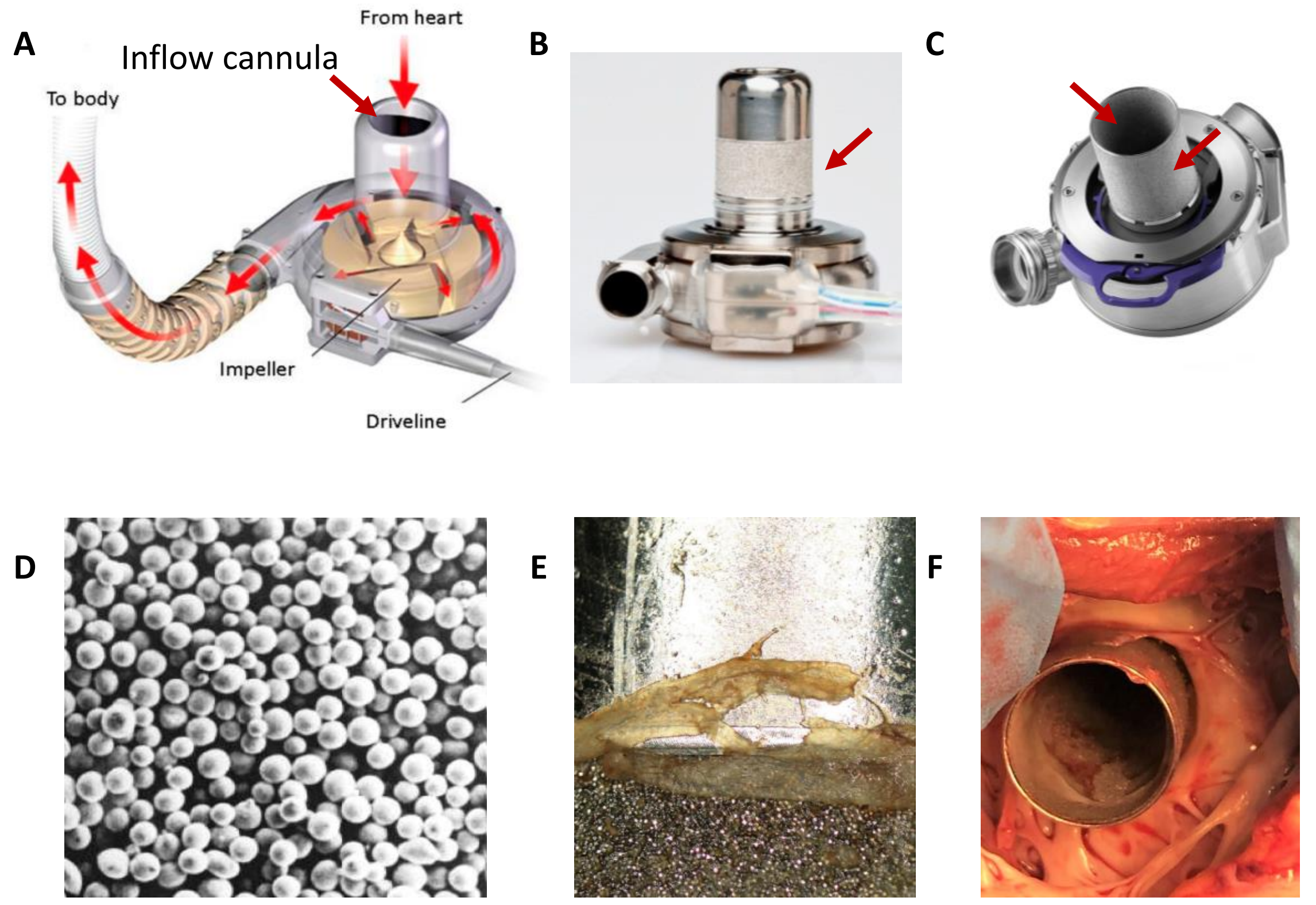

Figure 1. Sintered LVAD inflow cannula (arrow). (A) systemic figure of location of the inflow cannula respective to the centrifugal LVAD (HVAD) (B) the HVAD cannula (C) HM3. (D) Sintered microsphere 50x. (E) close-up view of thrombus on the HVAD cannula at the interface between the sintered and smoot region [8]. (F) Explanted HVAD cannula with deposition at the rim and sintered interior. Photo Courtesy of Dr. B. P. Griffith.

Blood Collection and Preparation

Anti-coagulated whole human blood (3.2% sodium citrate) was collected from 17 healthy donors (10 males, 7 females) at Cornell University's Human Metabolic Research Unit (HMRU, Ithaca, NY, USA). Written consent was obtained under an approved Institutional Review Board (IRB) protocol (approval date April 22, 2020). Participants were healthy, 18-65 in age, mean age of 25 $\pm$ 5 years, and abstained from anti-platelet drugs for at least 14 days. The first 2ml of blood collected was discarded in case of elevated activated platelets due to the venipuncture. Human whole blood was excluded with Hct, RBC, and platelet count below 35%, 4 x $10^{-6}$/mL, and 1.40 x $10^{-5}$/mL, respectively [20], which were measured with HMII hematology analyzer (Abaxis Inc., Union City, CA USA) [21]. Platelets were fluorescently stained with mepacrine through an established protocol [19] to visualize their aggregation and deposition during the study.

Microfluidic Channel Design

A custom microfluidic channel was designed for this study comprised of a laser-etched stainless steel chip (Great Lakes Engineering, Maple Grove, MN, the USA) sandwiched between a PDMS manifold and medical-grade silicone rubber sheet (BioPlexus, Kingman, AZ, USA) and glass slide, clamped between two acrylic plates. The flow path featured periodic teeth that emulated the sintered surface treatment of an inflow cannula. (See Figure 2.) The dimensions of the teeth were verified using a 3D laser scanning confocal microscope and accompanying software (Keyence Corp., Osaka, Japan). Samples were cleaned with Tergazyme® (Alconox Inc., New York, NY USA) and sonicated in distilled water under a previously established protocol before use [22].

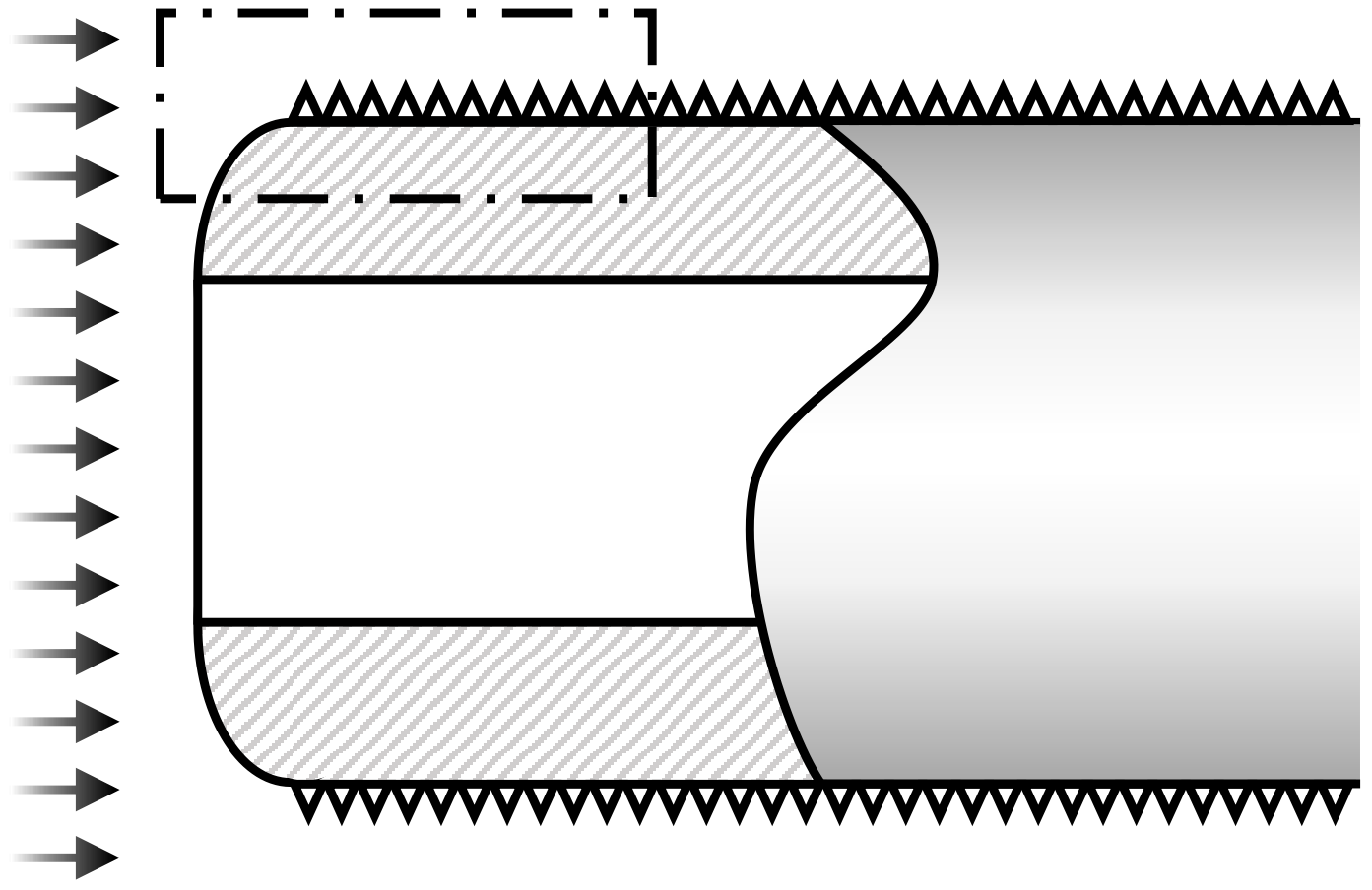

Figure 2. Cross-section of an LVAD cannula. Arrows indicate the blood flow direction. Region of interest indicated by rectangle. Emphasis on the rounded edge of the cannula rim and roughened surface along the exterior.

Blood Perfusion and Optics

A syringe pump (Harvard Apparatus, Holliston, MA USA) was used to perfuse blood through the microchannel at calibrated flow rates of 102, 252, 762 µL/ml, corresponding to Reynolds numbers of 3.06, 7.56, 22.88, to achieve shear rates of 1000, 2500, and 7500 $s^{-1}$, representative of values reported in the literature [23–25], which were confirmed both analytically and computationally. The three shear rates were selected because they are representative of VAD cannula at the bottom, in the middle and at the leading edge of the tip.

An inverted epifluorescence microscope (see Figure 3, Olympus Corp, Shinjuku, Japan) was assembled with a 103W HBO short arc mercury lamp (ASRAM GmbH, Munich, Germany) and CCD camera (Teledyne Photometrics, Tucson, AZ USA) to acquire images in real-time (7.5 FPS) through a 4x (W.D. 18.5 mm, NA 0.10; Olympus) or 10x (W.D.: 10mm, NA 0.30; Olympus) objective. The NIS software (Nikon) was used to acquire MP4 videos during blood perfusion.

Fluorescent Image Analysis

The ensemble-averaged platelet deposition was evaluated by the fluorescent intensity averaged over five (5) separate experiments, according to the equation **Error! Reference source not found.**

$$\bar{I}(x,y) = \frac{1}{N}\sum_{i}^{N} I_i(x,y) \quad (1)$$

where $i$ is $i$th image acquired for a particular shear-rate. Post-processing was performed by first extracting the region of interest and performing background subtraction using ImageJ (NIH, Bethesda, MD, USA). The frames were then aligned automatically using a custom code written in MATLAB (MathWorks, Inc., Natick, MA, USA) whereafter the quantities in **Error! Reference source not found.**, **Error! Reference source not** **found.**, and **Error! Reference source not found.** were computed. A comparison of simulation data against the ensemble average of the experimental data was performed.

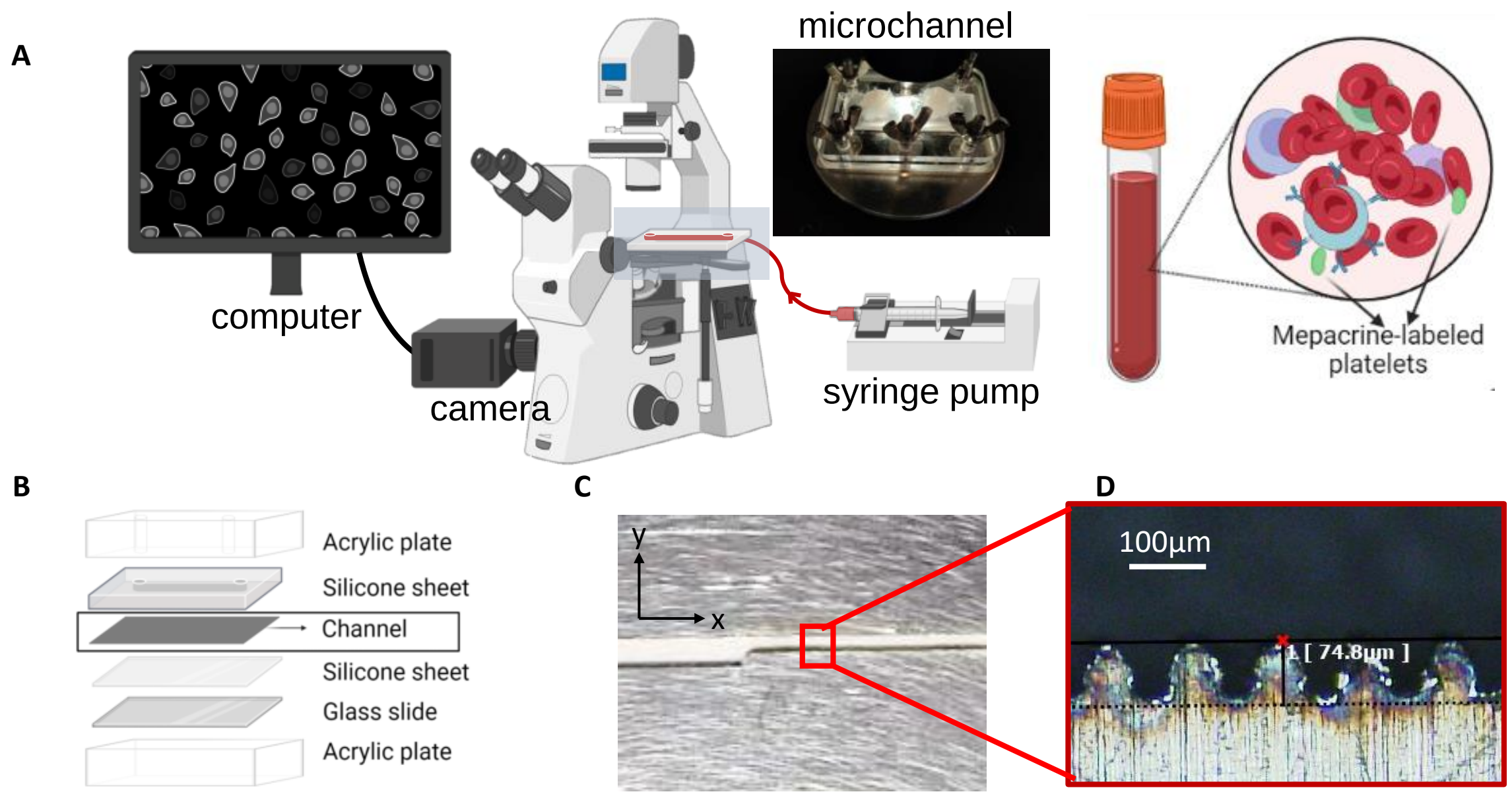

Figure 3. Experimental setup and microchannel design. (A) A representative image of CCD camera acquiring real-time images under an inverted epifluorescence microscope and blood perfusion into the microchannel (top left) utilizing a spring pump. The blood analog (right.) (B) The exploded view of the microfluidic device. (C) The stainless steel microchannel featuring periodic teeth that emulates the sintering surface of inflow cannulas. (D) Magnified view of serrated teeth under the laser profilometer.

Simulation of Thrombus Deposition

Numerical simulations were performed using the open-source CFD software OpenFOAM. The non-linear term in the momentum equation was discretized using a second-order upwind scheme. The temporal discretization scheme is the first-order Euler method. The numerical schemes were chosen such that robust convergence was achieved. The flow field was simulated in a 3D symmetric domain featuring a serrated surface following the CAD model of the designed microchannel. A hexahedral dominant mesh was used with two layers of prism elements bordering the boundary adjacent to the serrated metal surface. The mesh was refined using the snappyHexMesh utility of OpenFOAM. (See Figure 4.) A mesh convergence study was performed on steady-state velocity fields to achieve an optimal mesh density. (See supplementary S1.) The steady-state flow field was input as the initial condition of the thrombosis deposition simulation.

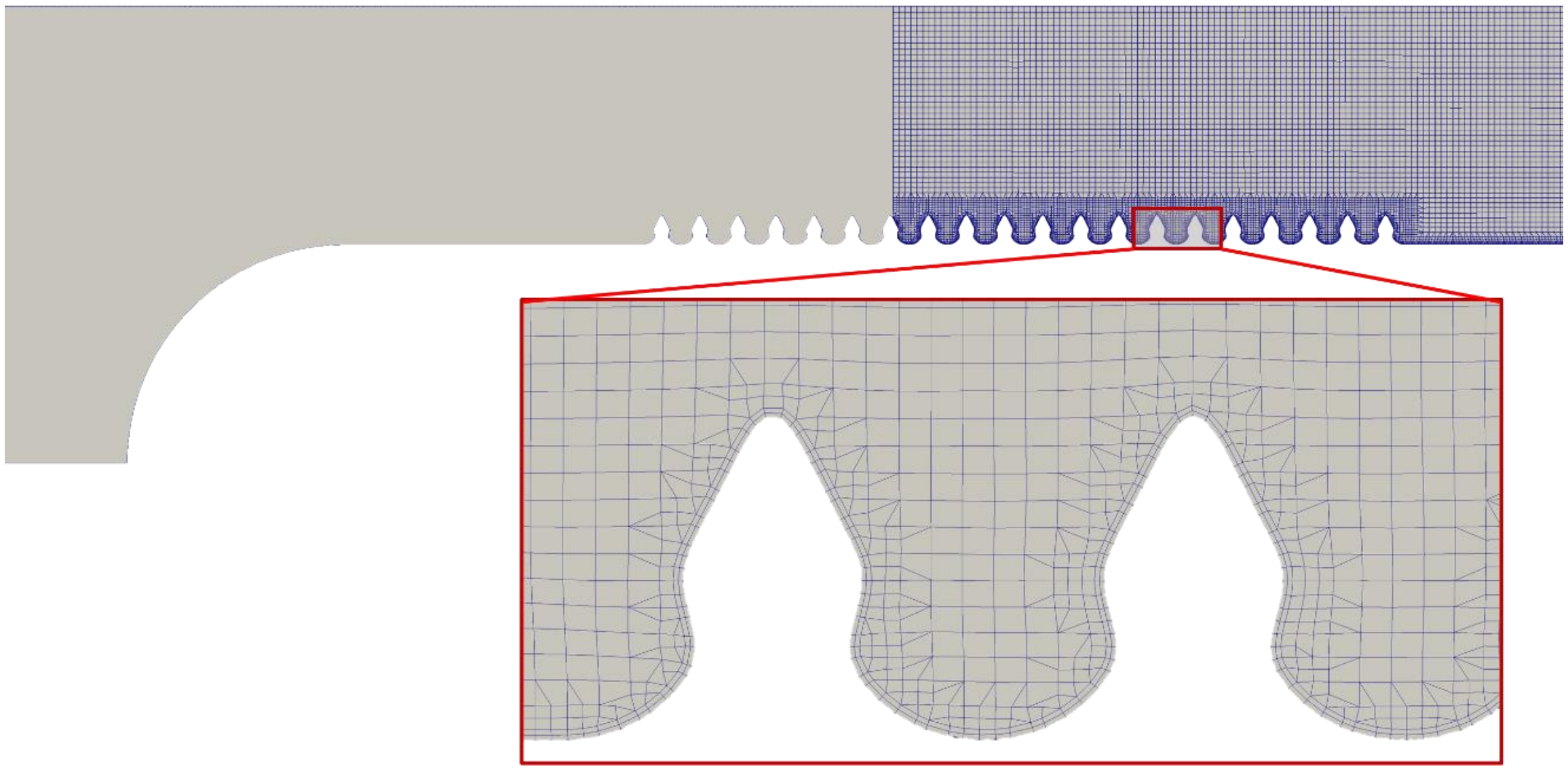

Figure 4. Computational domain encompassing the geometry(top) and mesh (top right) with refinement region(bottom).

Platelet deposition was simulated using a previously validated multi-constituent continuum model of thrombosis, implemented in OpenFOAM [26]. To account for the effect of surface chemistry, we used the calibrated rate of platelet deposition onto the bare metal [27] as the boundary condition. The remaining parameters of the model were taken from Zhussupbekov et al.[26].

One of the critical factors considered was von Willebrand factor (vWF) kinetics since extensional flow unfolds the vWF strands leading to vWF-platelet binding in high-shear regions [28]. Extensive literature supports that the presence or absence of vWF leads to distinct outcomes under the same supraphysiological hemodynamics [29,30].

To determine the relevance of vWF in this study, we performed a pilot *in vitro* investigation to visualize its presence in the serrated region under different shear rates (1000 $s^{-1}$, 2500 $s^{-1}$, 7500 $s^{-1}$, and 25000 $s^{-1}$) and bonding substrate (metal or collagen-covered metal). The 25000 $s^{-1}$ and collagen coated surface serve as a positive control in which anti-vWF conjugated antibody CY3 and TRITC filter are used (Chroma Tech. Corp, Bellows Falls, VT USA). The vWF staining results revealed co-locations of the vWF with the platelet deposition in the positive control group (Supplementary Figure S1) but not in the experiment group: 7500 $s^{-1}$ or below, nor in the bare metal. Therefore vWF-mediated thrombosis was excluded from the CFD simulation.

## Results

*In vitro* Platelet Deposition

The average fluorescence intensity map (Figure 5) depicts the time-course deposition on the serrated region at low, middle, and high perfusion rates corresponding to shear rates 1000 $s^{-1}$, 2500 $s^{-1}$, and 7500 $s^{-1}$, respectively. The blood flow direction is from left to right. The sustained growth of platelet adhesion with time was seen in all shear conditions suggesting the platelet deposition rate is greater than the embolization rate under all three shear rates for the duration of the study. Comparing results of various shear rates at the same perfusion time, low shear condition generated the greatest deposition at time points 1 min and 20 min. (See Figure 6) No significant difference in deposition was observed between the upstream and downstream locations.

Significant variances in deposition pattern were observed associated with the bulk flow shear rate. At a wall shear rate of 1000 $s^{-1}$, platelets were trapped in the valley between teeth primarily. Over time, the deposition gradually built up and expanded to the whole surface of the teeth. After 20 minutes of perfusion, the deposited platelets formed a confluent layer, covering the tips of the teeth to the bottom of the valleys. At the greatest shear, thrombi formation on the tips of teeth was less prevalent. This may be due to the elevated shear rate on the tips of the teeth causing a higher likelihood of embolization.

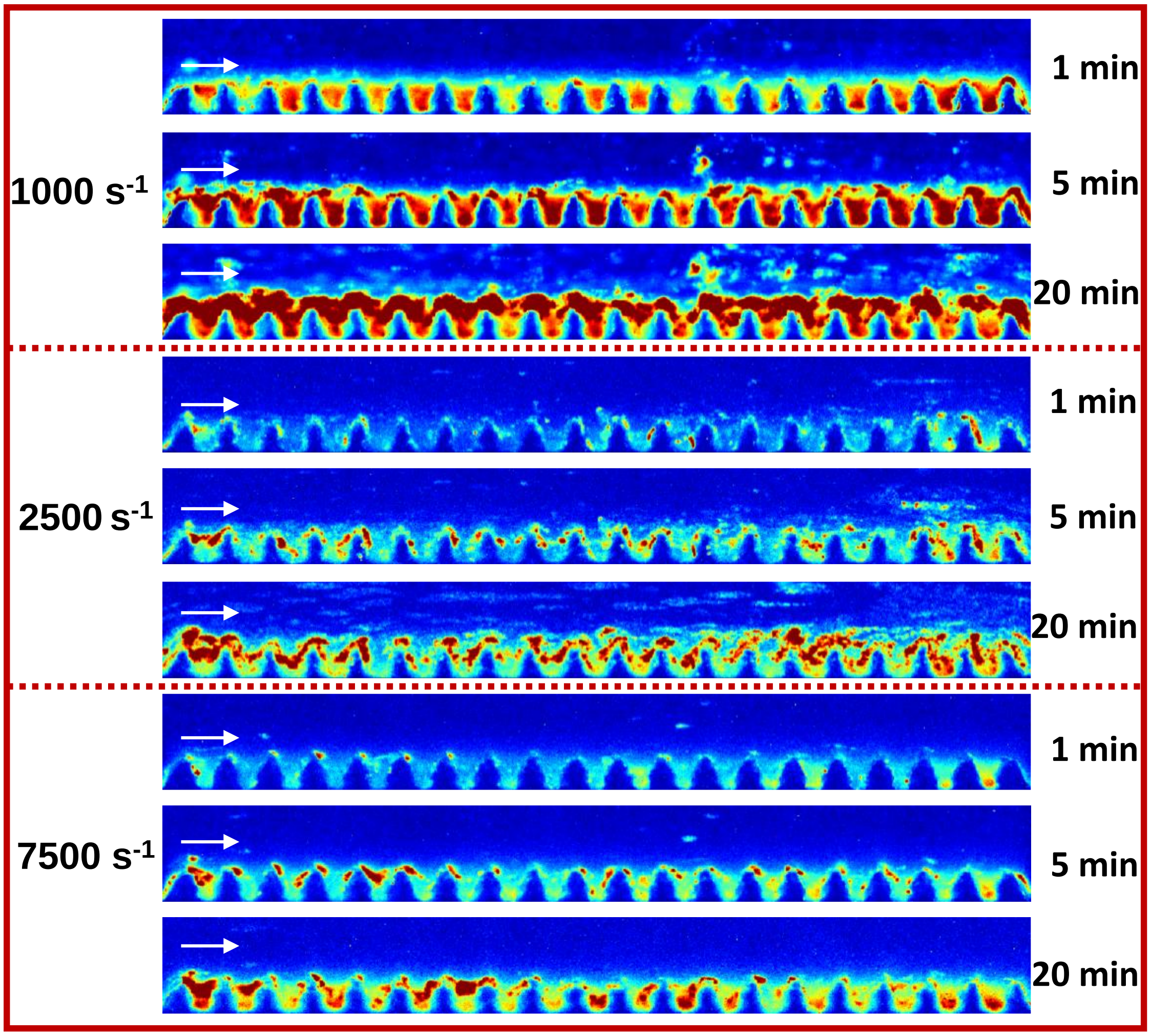

Figure 6. Time-course deposition on the serrated channel at low, middle, and high perfusion rates associated with shear rate 1000 $s^{-1}$, 2500 $s^{-1}$, and 7500 $s^{-1}$, respectively. From top to bottom are average deposition (n=5) intensity at 1 min, 5 min, and 20min. The flow is from left to right.

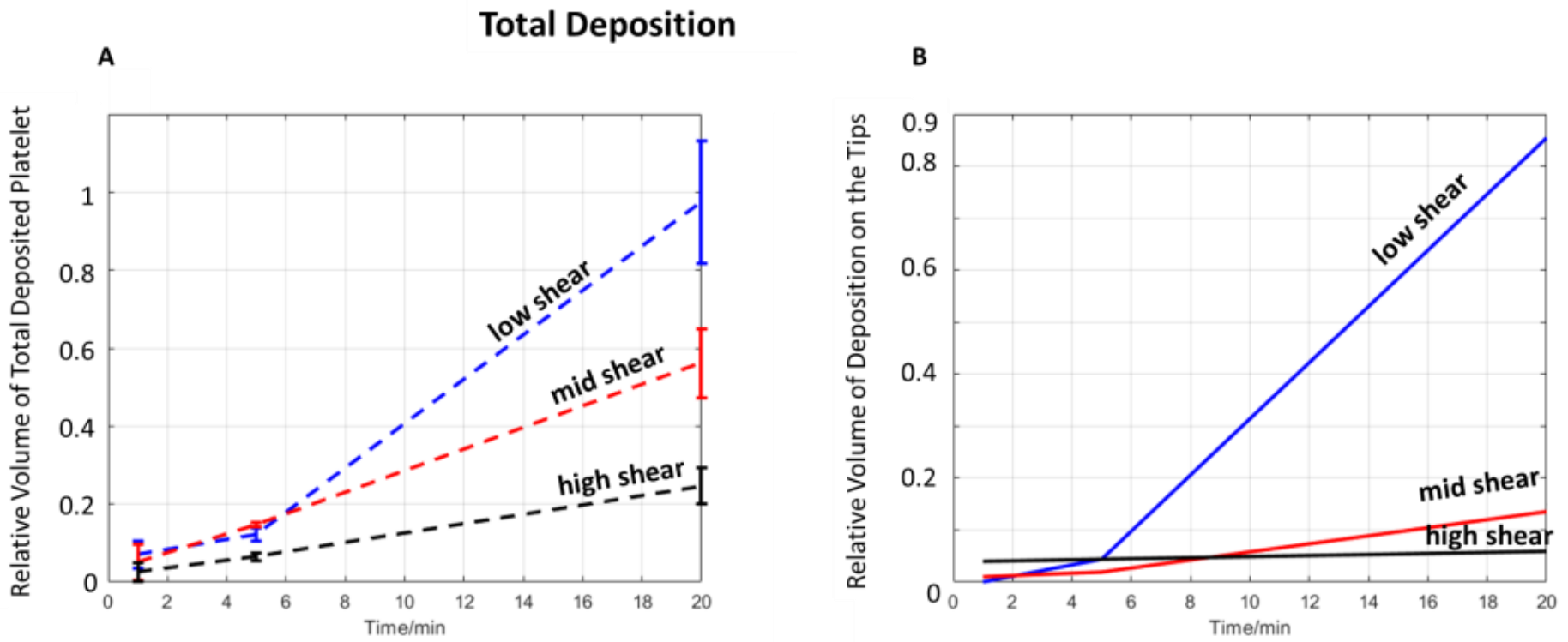

Figure 5. (A) Temporal average of total platelet deposition over n=5 experiments. The higher shear yields less deposition. (B) Temporal average of platelet deposition onto the tips of the saw-teeth (n=5). The residence time of deposited platelets at the tips reduces with increasing shear rate.

Computational Simulation

The computational simulation of the wall shear rates and flow fields within the serrated region is shown in Figure 7. The wall shear rate at the rounded leading edge was 1000 $s^{-1}$, 2500 $s^{-1}$, and 7500 $s^{-1}$ respectively. Regardless of the flow rate, the greatest wall shear rate is located at the tips of the teeth. The variance of deposition location was further confirmed by the *in silico* study in which the vWF pathway was deactivated (Figure 8). At 10-minute mark, it was observed that less shear rate resulted in more deposition on the tips of the teeth compared to high shear cases where deposition primarily occurs between teeth. This simulation result agrees with the *in vitro* platelet deposition map.

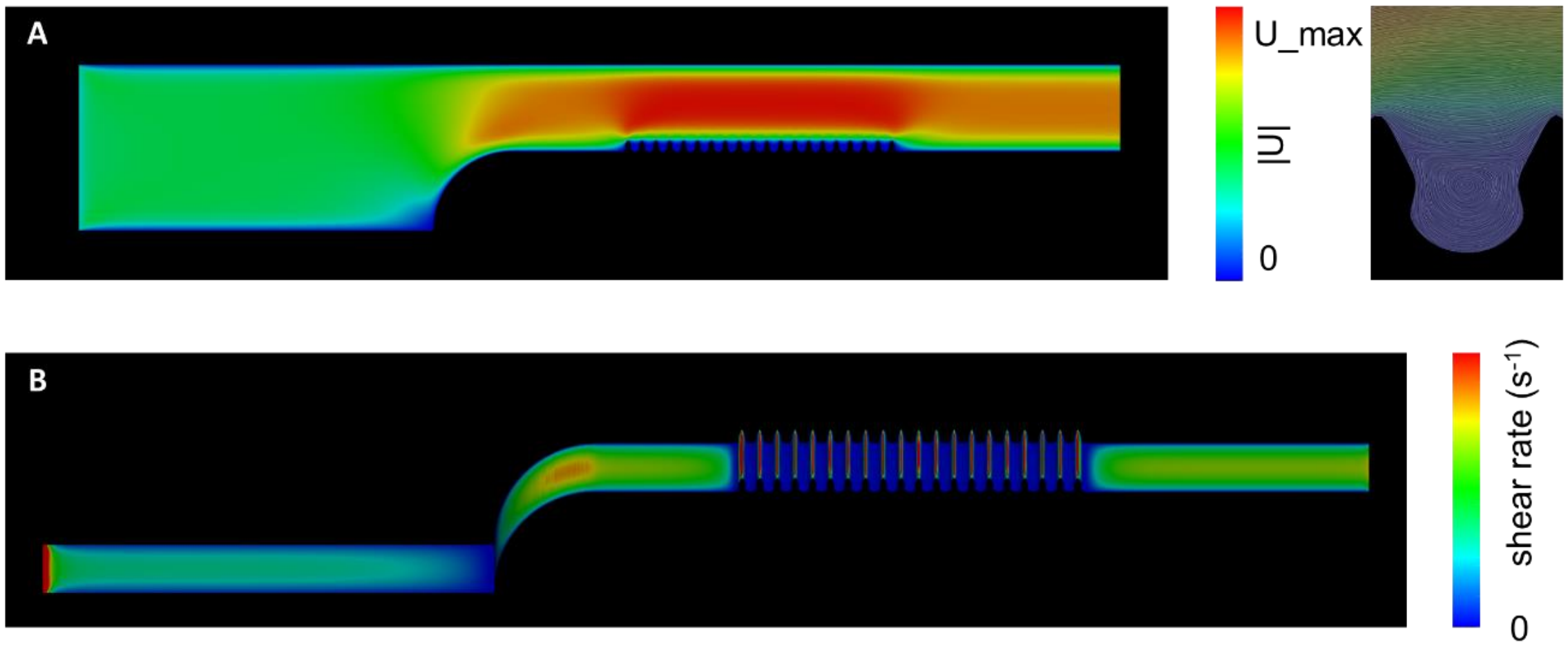

Figure 7. CFD simulation of the flow field in the channel and the wall shear rate at the bottom wall exhibits similarity across all three cases where the wall shear rate at the leading edge of the channel is 1000 $s^{-1}$, 2500 $s^{-1}$, 7500 $s^{-1}$. (A) The magnitude of velocity and streamline. The maximum velocity in the legend is 0.038 m/s, 0.095 m/s, and 0.29 m/s in low, middle, and high shear settings respectively. (B) Wall shear rate at the bottom wall with maximum legend of 1000 $s^{-1}$, 2500 $s^{-1}$, 7500 $s^{-1}$.

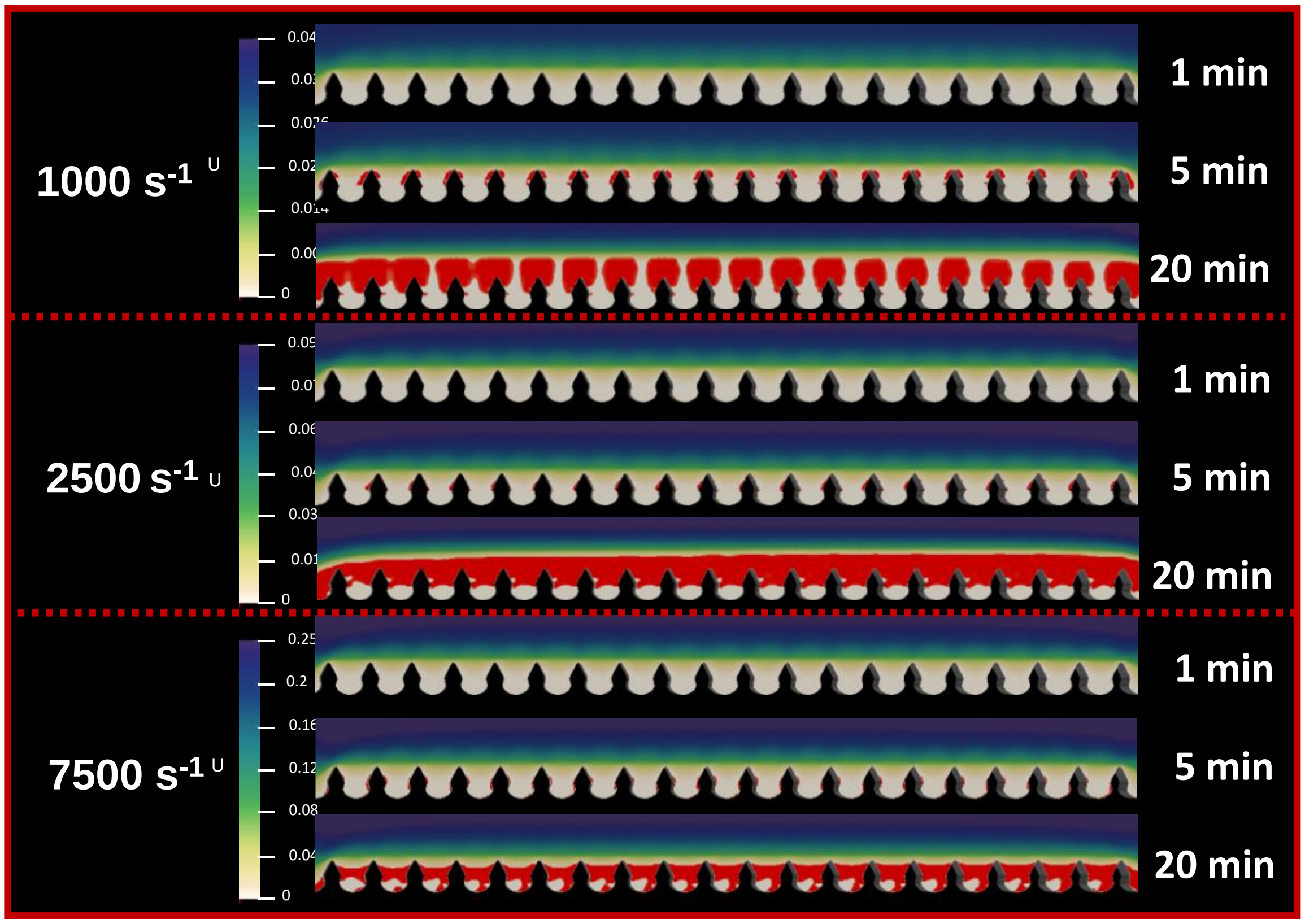

Figure 8. Thrombosis simulation within the channel at a perfusion rate of 102 µl/min and a wall shear rate of 1000 $s^{-1}$, at a perfusion rate of 252 µl/min and a wall shear rate of 2500 $s^{-1}$, and at a perfusion rate of 762 µl/min and a wall shear rate of 7500 $s^{-1}$.

**Discussion**

For decades, thrombotic stroke has persisted as a leading cause of morbidity and mortality among LVAD patients. The high incidence of stroke may be attributed to thrombi that form on the inflow cannula, which potentially embolize, are ingested, and pass unobstructed through the LVAD, because of the relatively large flow path, and potentially occlude an intracranial vessel [31,32]. Despite the continuing effort to develop a biocompatible and thrombo-resistant layer by incorporating sintered surfaces to the interior and exterior of the cannula, this approach has not yet been proven to be the panacea to effectively prevent stroke. In this study, we focused on the initial step of the biological response to the roughened surface.

Recent investigations have shown an increasing focus on surface roughness and shear as two collective triggers for thrombotic deposition [18,33–35]. Our study explored the higher range of shear rate (up to 7500 $s^{-1}$), unlike our previous study [19] which was focused on wall shear rates ranging from 400 $s^{-1}$ to 1000 $s^{-1}$. Both studies found a negative correlation between shear rate and platelet deposition [19]. This is contrasted to a recent study investigating *micro-scale* roughened Ti alloy, where platelet aggregation size was directly correlated to shear rate in the range from 1000 $s^{-1}$ to 5500$s^{-1}$ [18]. Although the deposition was observed to increase over the 20-minute duration (Figure 5), it is understood that accumulation is the difference between deposition and shear cleaning [13]:

$$\text{accumulation} = \text{deposition} - \text{clearance} \quad (2)$$

Therefore, the inverse correlation of deposition and shear can be explained by the increased rate of shear cleaning, exceeding the rate of deposition. It is likely that the deposition rate also increased with shear due to the increased adhesivity of shear-activated platelets. This phenomenon may explain the absence of thrombus deposition within the LVAD (Pump Thrombosis) [7] and the elevated risk of stroke. Platelet clearance, in turn, could occur by shedding of individual platelets, platelet microaggregates, and/or macro emboli. Our observation of the videographic recordings (see supplementary video S2) over the 20-minute duration of these studies revealed continuous sloughing of emboli in the range of 5-100 microns. (Individual platelets were not observable in our study.) Not surprisingly, the greatest rate of shedding occurs at the tip of the teeth (Figure 6) where the shear rate is much greater compared to the valleys between teeth. (See Figure 7.)

Our study aimed to emulate geometry-dependent hydrodynamic conditions around LVAD inflow cannula, and wall shear rates consistent with the previous study. Although the microchannel we studied was inspired by the sintered surface of the LVAD, we recognize that a limitation of this study is the disparity between the toothed surface of the microchannel and a sintered surface, comprised of microspheres.

In addition, it is important to note the disparity in material properties on thrombogenesis. Our study revealed time-course thrombosis on stainless steel, different from titanium alloy typically used in LVAD cannula. A previous study verified that stainless steel exhibits lower thrombogenicity compared to other biomaterials at low shear rates (150 $s^{-1}$), but accumulates more thrombi at shear rates higher than 5000 $s^{-1}$ which are in a laminated structure [33]. On the other hand, the titanium-rich oxide layer similar to that reported on titanium alloys improved the hemocompatibility of Niti which exhibits similar platelet adhesion patterns with stainless steel [36]. Given the disparity in thrombotic behavior between stainless steel and the Ti alloy, in the future, it is beneficial to incorporate a roughened Ti alloy surface that mimics the stochasticity of sintered cannula.

The continuum thrombosis model utilizes a shear-dependent platelet cleaning flux to account for embolization. The understanding of the mechanism and the accurate prediction of thrombosis are limited by the absence of an explicit model of embolization in our simulation model. In contrast to the low Reynolds numbers regime in our study, a very recent paper on the thrombo-embolization model employed turbulence modeling to reproduce embolization mechanics in a range of Reynolds numbers from 3000 to 16000 and corresponding shear rate 1200 $s^{-1}$ to 4400 $s^{-1}$ [37]. This highlights the necessity of our study as a baseline for investigating the adhesivity between thrombi and surfaces, going beyond shear stress considerations alone.

**Conclusion**

This study comprises *in vitro* and *in silico* counterparts, which emulate clinical-related surface roughness and supraphysiological shear of an LVAD inflow cannula. The *in vitro* visualization method provides a way to rigorously study the effect of supraphysiological shear to understand cannula thrombogenicity. Our *in vitro* observations highlight the importance of embolization while evaluating the hemocompatibility of a biomaterial. The *in silico* method predicts the cannula thrombus formation with variable flow conditions. Therefore, this study sheds light on flow conditions to improve hemocompatibility and hence to reduce the risk of stroke as well as motivates the development of comprehensive embolization physics in computational models. Continuing study to incorporate a roughened Ti alloy surface that mimics the stochasticity of the sintered cannula or optimizing the surface roughness of the cannula may be beneficial.

Future *in silico* models should consider incorporating embolization physics to develop a comprehensive understanding of the risk factors leading to stroke.

**Declaration of Generative AI and AI-assisted technologies in the writing process**

Statement: During the preparation of this work the author(s) used chatGPT solely for editing purposes to improve readability and language. After using this tool/service, the authors reviewed and edited the content as needed and take full responsibility for the content of the publication.

**Data Availability Statement**

The raw data required to reproduce these findings are available to download from [https://www.dropbox.com/scl/fo/7lu0b8jf0aci5imdqg6di/h?rlkey=2s58fizxqfrjkgsf11kjhs69a&dl=0]. The processed data required to reproduce these findings are available to download from [https://www.dropbox.com/scl/fo/9v2gsvhxfbwpvgciu5irf/h?rlkey=lxt7r23ok1uun1hn2yef3h7qt&dl=0]

**Acknowledgments**

We would like to extend our gratitude to the National Institute of Health Grant HL089456 for funding this study.